\documentclass[twocolumn,prl,aps,showpacs,floatfix,superscriptaddress]{revtex4}
\begin{document}

\noindent {\bf  Comment on ``Trouble with the Lorentz Law of Force: Incompatibility with Special Relativity and Momentum Conservation''}

\vspace{0.3cm}

% Summary
In a recent Letter \cite{MansuripurPRL} Mansuripur considers 
a magnetic point dipole $m_0 \, \hat {\bf x'}$ positioned in the rest frame at a fixed location $d \, \hat {\bf z'}$ from a point charge $q$. 
Performing a Lorentz transformation to a laboratory frame
where the charge distribution moves at velocity $V \, \hat {\bf z}$ he finds that 
`a net torque ${\bf T}=(V q m_0/ 4 \pi d^2)\, \hat {\bf x}$ acts on the dipole pair'.
He then argues that `this torque in the (lab) frame in the
absence of a corresponding torque in the (rest) frame is sufficient proof of the 
inadequacy of the Lorentz (force) law'.

%Solution of paradox
In this comment we demonstrate that the presence of a torque 
is {\it not} incompatible with special relativity and momentum conservation.
In fact, the torque is {\it required} by the conservation
laws that apply to the {\it total} momentum of the system (including the particles).
%General statments on classical em incomplete
We furthermore stress that classical electrodynamics
needs a consistent dynamical description of the particles,
a point not addressed in \cite{MansuripurPRL}.
This description involves coupled equations for the electromagnetic fields and the 
trajectories \cite{BrachetTirapegui78} and any conserved quantity will then contain 
contributions from both the field and the particles.

%General conservation laws
The standard momentum conservation reads in $4D$ relativistic notations 
$\partial_\alpha (T_{\rm particle}^{\alpha \beta}+T_{\rm field}^{\alpha \beta})=0$, 
with $\partial_\alpha T_{\rm field}^{\alpha \beta}= -c^{-1} F^{\beta \lambda} j_\lambda$ \cite{Jackson}.
The latter implies the $3D$ relation
$\partial_t u + \nabla \cdot {\bf S} = - {\bf E} \cdot {\bf J_{\rm total}}$
 with $u=\epsilon_0({\bf E}^2 +c^2{\bf B}^2)/2$ and ${\bf S}=\mu_0^{-1}{\bf E} \times  {\bf B}$.
Poynting's theorem yields the total field momentum $\mathcal{P}_{\rm field}=c^{-2}\int {\bf S} \, d^3 x$ and 
angular momentum $\mathcal{L}_{\rm field}=c^{-2}\int  ({\bf r}-{\bf r}_0) \times {\bf S} \, d^3x$.

%Computation Dipole
It is straightforward to find the momenta in the rest frame by direct (but tedious) computation. 
A faster derivation is obtained by
considering an auxiliary problem \cite{DipolesMomentum} involving an extra $\delta$ distributed contribution to ${\bf B}$ 
and $2$ charge monopole \cite{DiracMonopole} pairs that are known to each have a 
total angular moment \cite{Jackson} $q g/4 \pi$ \footnote{The Dirac quantization condition corresponds to this 
angular momentum being an integer multiple of $\hbar/2$.}. The momenta in the rest frame read
\begin{eqnarray}
\mathcal{P}_{\rm field}&=&\frac {1} {4 \pi } \frac{m_0 q} {d^2} \hat {\bf y'}\label{Eq:defP}\\
\mathcal{L}_{\rm field}&=& \frac {1} {4 \pi } \frac{m_0 q} {d} \hat {\bf x'}+(d \, \hat {\bf z'}-{\bf r}_0') \times \mathcal{P}_{\rm field}
\label{Eq:defL}
\end{eqnarray}

%Result
The standard momentum conservation law $\partial_\alpha T_{\rm particle}^{\alpha \beta}= c^{-1} F^{\beta \lambda} j_\lambda$
in a time-independant situation requires that $\mathcal{P}_{\rm particle}+\mathcal{P}_{\rm field}=0$. This momentum 
$\mathcal{P}_{\rm particle}=-({m_0 q}/ {d^2}{4 \pi }) \hat {\bf y'}$ corresponds to the notion  of `missing momentum' 
\cite{ShockleyJames,vaidman90}
and the associated position dependent contribution to $\mathcal{L}_{\rm total}$ cancels the last term of (\ref{Eq:defL}).
In the lab frame the presence of the torque is required to account for the motion at uniform speed of the missing momentum $\mathcal{P}_{\rm particle}$.

%works alike for free and bound currents
A simple thought experiment shows the unavoidability of the missing momentum. Consider, instead of the magnetic point dipole, an electromagnet that is smoothly turned off at $t=0$ with negligible radiative effects. Using $\nabla \times {\mathbf E}=-\partial{\mathbf B}/\partial t$ it is straightforward to compute (masses being large enough to neglect motions) that the charge picks up a momentum of magnitude $\mathcal{P}_{\rm field}$. The missing momentum $\mathcal{P}_{\rm particle}$ allows for the corresponding recoil of the electromagnet that is needed for linear and angular momentum conservation \cite{ShockleyJames}. Note that this argument is independent of the nature (possibly containing magnetic material such as soft iron) of the electromagnet.

%no magic bullet!
Let us finally stress that the fundamental incompleteness of 
Maxwell's equations and the Lorentz force law (when they are taken alone)
is not addressed in \cite{MansuripurPRL}. 
What is lacking is a dynamical description of the charged particles
response to the Lorentz force consistent with the fact that particles radiate when their accelerations vary.
%only way to complete it at this level is lorentz dirac
The standard answer to this classical problem is the Lorentz-Dirac equation \cite{DiracLorentz}.
To obtain this fully coupled description
a cut-off-dependent (divergent but unobservable) renormalization of the 
mass of the particle has to be performed. 
%Lagrangian gives lorentz force, etc.
The Lorentz-Dirac equation 
is known \cite{BrachetTirapegui78} to
generally describe
the interaction of a classical field with a particle, independently of the details 
of the interaction, provided that the coupled equations of motion are obtained 
from a Lagrangian.

E. T.  acknowledges support from FONDECYT Project 1120329.

\vspace{0.2cm}

\noindent {M. Brachet} $^1$ {and}  {E. Tirapegui} $^2$

\noindent$^1${Laboratoire de Physique Statistique de l'Ecole Normale Sup{\'e}rieure,
associ{\'e} au CNRS et aux Universit{\'e}s Paris VI et VII,
24 Rue Lhomond, 75231 Paris, France}

\noindent$^2${Departamento de F\'\i sica, Facultad de Ciencias F\'\i
sicas y Matem\'aticas de la Universidad de Chile, Blanco Encalada
2008, Santiago, Chile.}

\vspace{0.2cm}


%\bibliographystyle{unsrt}
%\bibliography{bibli}

\begin{thebibliography}{8}
\expandafter\ifx\csname natexlab\endcsname\relax\def\natexlab#1{#1}\fi
\expandafter\ifx\csname bibnamefont\endcsname\relax
  \def\bibnamefont#1{#1}\fi
\expandafter\ifx\csname bibfnamefont\endcsname\relax
  \def\bibfnamefont#1{#1}\fi
\expandafter\ifx\csname citenamefont\endcsname\relax
  \def\citenamefont#1{#1}\fi
\expandafter\ifx\csname url\endcsname\relax
  \def\url#1{\texttt{#1}}\fi
\expandafter\ifx\csname urlprefix\endcsname\relax\def\urlprefix{URL }\fi
\providecommand{\bibinfo}[2]{#2}
\providecommand{\eprint}[2][]{\url{#2}}

\bibitem[{\citenamefont{Mansuripur}(2012)}]{MansuripurPRL}
\bibinfo{author}{\bibfnamefont{M.}~\bibnamefont{Mansuripur}},
  \bibinfo{journal}{Phys. Rev. Lett.} \textbf{\bibinfo{volume}{108}},
  \bibinfo{pages}{193901} (\bibinfo{year}{2012}).

\bibitem[{\citenamefont{Brachet and Tirapegui}(1978)}]{BrachetTirapegui78}
\bibinfo{author}{\bibfnamefont{M.}~\bibnamefont{Brachet}} \bibnamefont{and}
  \bibinfo{author}{\bibfnamefont{E.}~\bibnamefont{Tirapegui}},
  \bibinfo{journal}{Nuovo Cimento Soc. Ital. Fis. A}
  \textbf{\bibinfo{volume}{47}}, \bibinfo{pages}{210} (\bibinfo{year}{1978}).

\bibitem[{\citenamefont{Jackson}(1998)}]{Jackson}
\bibinfo{author}{\bibfnamefont{J.~D.} \bibnamefont{Jackson}},
  \emph{\bibinfo{title}{{Classical Electrodynamics}}}
  (\bibinfo{publisher}{Wiley}, \bibinfo{year}{1998}), \bibinfo{edition}{3rd}
  ed.

\bibitem[{\citenamefont{{Griffiths}}(1992)}]{DipolesMomentum}
\bibinfo{author}{\bibfnamefont{D.~J.} \bibnamefont{{Griffiths}}},
  \bibinfo{journal}{Am. J. Phys.} \textbf{\bibinfo{volume}{60}},
  \bibinfo{pages}{979} (\bibinfo{year}{1992}).

\bibitem[{\citenamefont{Dirac}(1931)}]{DiracMonopole}
\bibinfo{author}{\bibfnamefont{P.~A.~M.} \bibnamefont{Dirac}},
  \bibinfo{journal}{Proc. R. Soc. A} \textbf{\bibinfo{volume}{133}},
  \bibinfo{pages}{60} (\bibinfo{year}{1931}).

\bibitem[{\citenamefont{Shockley and James}(1967)}]{ShockleyJames}
\bibinfo{author}{\bibfnamefont{W.}~\bibnamefont{Shockley}} \bibnamefont{and}
  \bibinfo{author}{\bibfnamefont{R.~P.} \bibnamefont{James}},
  \bibinfo{journal}{Phys. Rev. Lett.} \textbf{\bibinfo{volume}{18}},
  \bibinfo{pages}{876} (\bibinfo{year}{1967}).

\bibitem[{\citenamefont{Vaidman}(1990)}]{vaidman90}
\bibinfo{author}{\bibfnamefont{L.}~\bibnamefont{Vaidman}},
  \bibinfo{journal}{Am. J. Phys.} \textbf{\bibinfo{volume}{58}},
  \bibinfo{pages}{978} (\bibinfo{year}{1990}).

\bibitem[{\citenamefont{Dirac}(1938)}]{DiracLorentz}
\bibinfo{author}{\bibfnamefont{P.~A.~M.} \bibnamefont{Dirac}},
  \bibinfo{journal}{Proc. R. Soc. A} \textbf{\bibinfo{volume}{167}},
  \bibinfo{pages}{148} (\bibinfo{year}{1938}).

\end{thebibliography}

\end{document}